# Multidimensionality of Legal Singularity: Parametric Analysis and the Autonomous Levels of AI Legal Reasoning

**Dr. Lance B. Eliot**
Chief AI Scientist, Techbruim; Fellow, CodeX: Stanford Center for Legal Informatics
Stanford, California, USA

## Abstract

Legal scholars have in the last several years embarked upon an ongoing discussion and debate over a potential Legal Singularity that might someday occur, involving a variant or law-domain offshoot leveraged from the Artificial Intelligence (AI) realm amid its many decades of deliberations about an overarching and generalized technological singularity (referred to classically as *The Singularity*). This paper examines the postulated Legal Singularity and proffers that such AI and Law cogitations can be enriched by these three facets addressed herein: (1) dovetail additionally salient considerations of *The Singularity* into the Legal Singularity, (2) make use of an in-depth and innovative multidimensional parametric analysis of the Legal Singularity as posited in this paper, and (3) align and unify the Legal Singularity with the Levels of Autonomy (LoA) associated with AI Legal Reasoning (AILR) as propounded in this paper.

**Keywords:** AI, artificial intelligence, autonomy, autonomous levels, legal reasoning, law, lawyers, practice of law, The Singularity, Legal Singularity

## 1 Background of The Singularity

In section 1 of the paper, the topic of *The Singularity* is introduced and addressed. Doing so establishes the groundwork for section 2, covering a form of singularity that has come to be known as the Legal Singularity (LS), considered to be an offshoot or a domain-specific variant of the overarching *The Singularity*. Section 3 indicates the Levels of Autonomy (LoA) of AI Legal Reasoning (AILR), which will be instrumental in the discussions undertaken in Section 4. Section 4 then provides an in-depth analysis of the Legal Singularity as it relates to the LoA of AI Legal Reasoning and lays out an important parametric analysis of the Legal Singularity. The final section, Section 5, covers additional considerations and future research.

This paper then consists of these five sections:

- Section 1: Background of The Singularity
- Section 2: Legal Singularity
- Section 3: Autonomous Levels of AI Legal Reasoning
- Section 4: Legal Singularity Multidimensionality, Alignment with LoA AILR
- Section 5: Additional Considerations and Future Research

Since the word "singularity" is used in at least two contexts within this paper, one context being an overarching or grandiose kind of singularity, typically referred to as *The Singularity*, and the other being a singularity specific to the field of law, known as the Legal Singularity, the convention in this paper will be that whenever referring to the Legal Singularity this will be done by stating "Legal Singularity" or by the abbreviation of "LS," while the larger *The Singularity* will be referred to as the "singularity" or "AI singularity" or "Technological singularity," and when desiring to especially emphasize such a reference it will be stated as *The Singularity* (such an emphasize is done solely for drawing attention to the matter and not due to suggesting any differences of meaning or connotation).

### 1.1 Understanding *The Singularity*

A longstanding discussion and debate in the field of Artificial Intelligence entails a controversy referred to as *The Singularity* [9] [14] [28] [39]. Sometimes also coined as the AI Singularity or the Technological Singularity, the concept underlying the matter is relatively ill-defined and has substantively varied in details of its meaning and substance over the now many years of its postulation (dating back to the 1950s).

Often first-traced to commentary by the famous mathematician and pioneering computer scientist John von Neumann, here is what researcher Ulman [58] in 1958 indicated had occurred in a conversation with Von Neumann: "One conversation centered on the ever-accelerating progress of technology and changes in the mode of human life, which gives the appearance of approaching some essential singularity in the history of the race beyond which human affairs, as we know them, could not continue."

Essentially, the sentiment at the time was that computers might eventually be able to achieve human intelligence, potentially even eclipsing human intelligence, and the result could be problematic for humanity. Of course, similar exhortations have been replete in science fiction, though typically proffered by imaginative writers with unsupported visions rather than by bona fide scientists that are making such speculations based on their assessment of the underlying technology and attempting to anticipate future outcomes. That's not to suggest that those scientists will necessarily be on par with predicting the future, and there are many documented instances of scientists that were wildly off-the-target and baseless in their prophesizing. In short, expertise in a subject matter is a worthwhile basis for providing meaningful predictions, nonetheless, that expertise can still be misguided or mistaken as to what the future might hold.

In 1965, Oxford researcher Irving John Good [36] published a cornerstone research paper that extended the singularity notion and tied the topic to the emergence of computers that might be considered as ultra-intelligent, commonly today referred to as aiming to be super-intelligent or super-human in capability. Rather than emphasizing the dangers associated with mankind developing an ultra-intelligent machine, Good [36] urged that the survival of humanity depended on being able to craft such a system and indeed ought to be done as soon as possible: "The survival of man depends on the early construction of an ultra-intelligent machine."

Here is Good's [36] definition associated with the capabilities envisaged: "Let an ultra-intelligent machine be defined as a machine that can far surpass all the intellectual activities of any man however clever. Since the design of machines is one of these intellectual activities, an ultra-intelligent machine could design even better machines; there would then unquestionably be an intelligence explosion, and the intelligence of man would be left far behind. Thus, the first ultra-intelligent machine is the last invention that man need ever make."

In this initial elucidation of the topic by Good, a key facet that has become inextricably woven into the singularity rubric is the speculative idea of an intelligence explosion. In short, if mankind can craft an AI system to some threshold of intelligence, it is presumed that the AI could then further progress, essentially on its own accord, by using its core base of intelligence to further produce more intelligence. No one as yet knows what this minimum threshold might be, and nor is there any viable means to anticipate how far the presumed intelligence explosion might proceed in terms of the upper limits of some unknown super-intelligence, raising the perennial question of how high is up, as it were.

Pursuing for the moment the somewhat tangential but relevant question concerning the notion of an intelligence explosion, let's consider the ramifications of such a phenomenon, if indeed possible (no one knows whether it is or not). Similar to questions that arose during the creation of the atom bomb, whereby scientists were somberly worried that the ignition and exploding of an atomic bomb might somehow catch hold and violently and rapidly spread across the globe in an unheralded conflagration, some assert that the same might happen in the case of an intelligence explosion. To wit, the intelligence being produced might magnify and expand, for which the result could be to have humans seem like mere ants in intelligence versus the super-intelligence spawned by AI. Again, some view this outcome as disastrous for humanity, possibly being enslaved by a super-intelligent AI, while others believe that mankind might be saved due to an artificial super-intelligence that could solve the gravest problems confronting the survivability of humans.

In his research, Good attempted to outline some of the overall features or elements that seemed at the time to potentially be required to achieve an AI super-intelligence. For example, he debated those theories of the period concerning a tremendous amount of parallelism that would be needed, a facet of modern-day computers that were not especially viable when Good [36] wrote his paper in the 1960s: "It cannot be regarded as entirely certain that an ultra-intelligent machine would need to be ultraparallel since the number of binary operations per second performed by the brain might be far greater than is necessary for a computer made of reliable components. Neurons are not fully reliable; for example, they do not all last a lifetime; yet the brain is extremely efficient. This efficiency must depend partly on 'redundancy' in the sense in which the term is used in information theory. A machine made of reliable components would have an advantage, and it seems just possible that ultraparallel working will not be essential. But there is a great waste in having only a small proportion of the components of a machine active at any one time."

At this time, there is still no definitive means to specify what the AI might be composed of that would lead to the singularity, and the work on these facets remains exploratory and speculative in nature. Furthermore, not everyone conceives of the singularity as necessarily requiring super-intelligence, and nor an intelligence explosion. Some proffer that the singularity might be considered as the reaching of human intelligence via AI, often referred to as Artificial General Intelligence (AGI). At that achievement alone, the singularity will have been achieved, some contend. Whether this then leads to a subsequent intelligence explosion, and some kind of super-intelligence, can be considered a separate aspect, for which the singularity then perhaps is essentially a furtherance or extension rather than an initial arrival per se.

Consider these variations:

a) Singularity is the achievement of AI such that the AI has attained human intelligence capabilities
b) Singularity is the eclipsing of human intelligence by AI and attaining a super-intelligence
c) Singularity is an intelligence explosion whereby AI generates or spawns further intelligence.

Arguments ensue as to which of those is the "singularity" and also whether they must be combined or co-existent to count as the singularity occurring. Perhaps "a" or arrival at human intelligence is mandatory for getting to "b," though others contend that it is possible that an AI decidedly less-than-human intelligence levels might percolate via "c" per an intelligence explosion and then nearly instantaneously exceed "a" and arrive at "b," thus never especially settling down at the mere capacity of human intelligence. Others argue that there is not anything feasibly beyond human intellectual capacities, regardless of how adept the AI might be, and as such the achievement of human intelligence is the capstone limit. In that case, the singularity would be solely about the "a" and not take into account the "b" and "c" postulates which are deemed as impossible and a false aspiration.

Additionally, some assert that a super-intelligence might be reached without any need for and indeed no occurrence at all of an intelligence explosion. In this claim, an intelligence explosion is unlikely, perhaps even impossible per se, and that the super-intelligence might arise is some other more "mundane" manner and not due to a speculative and seeming spectacular intelligence explosion. Variations of that theory are that an intelligence explosion might very well occur, but it will be more of a whimper than a bang, such that the explosion is not particularly explosive. Instead, envision that intelligence oozes or synergizes with other intelligence, doing so demurely, rather than in a highly combustive manner.

The variants consist of these possibilities (but not limited to) that might constitute *The Singularity*:

- "a" only
- "b" only
- "c" only
- "a" and "b" mandatory
- "b" and "c" mandatory
- "a" and "c" mandatory
- "a" and "b" and "c" mandatory
- Other

For purposes of this discussion, we relax the potential requirement that the singularity must have a super-intelligence and also that an intelligence explosion is required. Since the singularity is already a highly conceptualized theory, to begin with, it seems reasonable to accept the notion that if human intelligence is achieved via AI, this accomplishment by itself would present the same potentiality as would the super-intelligence and the intelligence explosion. In essence, wherein some believe that only a super-intelligence bodes for the outcomes of either goodness or badness for society, there is ample room to equally speculate that an AI of human intelligence capability could likewise give rise to magnitudes of goodness and badness for society. Keep in mind that such an AI would presumably be able to be replicated and shared all across the globe, giving a boost to its potential impacts, regardless that it might not be deemed as super-intelligent or was not borne from an intelligence explosion.

It is further conceivable that we might split the difference and find a middle ground of a singularity at phase one, consisting of reaching human intelligence, and then a phase two of singularity that involves super-intelligence. Some are not satisfied with this two-phased division and insist that the singularity can only occur if super-intelligence is achieved. This is somewhat a disingenuous contention though if there is no means to codify or specify what the super-intelligence consists of. Indeed, one of the ongoing disputes about the very notion of super-intelligence is that there is a paucity of substantive criteria to pin down the capacities that super-intelligence presumably guarantees, raising the question of how will we know that super-intelligence has been attained and that it is not an AI-based human level of intelligence that just so happens to seem super-intelligent from our perspective. Those that proffer the archetypal "I'll know it when I see it" retort are not especially contributory to these serious-minded deliberations.

The renowned book about AI Singularity by Ray Kurzweil [39] in 2006 has become a foundational treatise on the topic of *The Singularity* and wrestles comparably with many of the potential facets of what the singularity constitutes. Kurzweil directly tackles the numerous criticisms about the singularity, including instances of doubt expressed by the assumed limits on neural processing, the Church-Turing thesis, Searle's Chinese Room Argument, theism, holism, and the like. After examining those numerous and varied critiques, Kurzweil [39] steadfastly asserts: "The Singularity, as we have discussed it in this book, does not achieve infinite levels of computation, memory, or any other measurable attribute. But it certainly achieves vast levels of all of these qualities, including intelligence. With the reverse engineering of the human brain, we will be able to apply the parallel, self-organizing, chaotic algorithms of human intelligence to enormously powerful computational substrates."

In that sense, it does not seem well-intentioned to debate in any preoccupied manner on the merits of AI Singularity as to whether a super-intelligence is attained, and nor whether there is an intelligence explosion, and instead concentrate on the overwhelming and overarching factor of AI embodiment of human-level intelligence, for which the potential outcomes are amplified when considering the presumed likelihood that this means that the AI could be readily replicated and distributed, doing so in a manner and form that heretofore of human intelligence could not be equally realized.

Kurzweil's book is provocatively titled as indicating that the singularity is near [39]. Others such as Walsh [61] offer a less optimistic timeline, indicating that the singularity is not only not near, it might not ever be near. Braga [9] suggests that despite whatever timing might be involved, the possibility of the singularity is surrounded by fallacies and that the debates about the singularity ought to be leveraged for considering how the dispute themselves gives rise to potential new opportunities in AI.

For speculations about the timing of the singularity, there are said to be potential "defeaters" that could undermine the postulated timelines. For example, many in the AI field are apt to offer that extraordinary and undefined AI breakthroughs have to occur if the singularity is going to be attained and that the timelines oft-stated are based on as-yet discovered technological innovations [58] [59]. Assumptions are made that technological progress on AI is going to proceed in some determinable fashion, and as long as that estimated path continues, the timing for the singularity remains on-target. Chalmers [14] offers this pronouncement about the role and nature of defeaters in this manner: "As for defeaters: I will stipulate that these are anything that prevents intelligent systems (human or artificial) from manifesting their capacities to create intelligent systems. Potential defeaters include disasters, disinclination, and active prevention. For example, a nuclear war might set back our technological capacity enormously, or we (or our successors) might decide that a singularity would be a bad thing and prevent research that could bring it about. I do not think considerations internal to artificial intelligence can exclude these possibilities, although we might argue on other grounds about how likely they are."

Another salient point frequently discussed about *The Singularity* involves the so-called Singularity Paradox. This is a presumed paradox that seems to undercut the doomsday scenarios that have been prophesied about the singularity. For example, one of the most famous doomsday indications involves the Paperclip Problem as described by Bostrom [8]. In this invented scenario, AI that has achieved singularity is asked by humanity to undertake the production of paperclips. The AI proceeds to do so, and takes this goal to an extreme, ultimately consuming all of the earth's production capacity to make paperclips. In the exceptionally sorrowful versions of the Paperclip Problem, the AI determines that humans are preventing the AI from maximizing the making of paperclips and thus does away with humans entirely. Though this might seem like a reasoned concern, Ympolskiy [65] explains why the Singularity Paradox is a worthwhile consideration to refute some of these doomsday manifestations: "Investigators concerned with the existential risks posed to humankind by the appearance of superintelligence often describe what we shall call a Singularity Paradox (SP) as their main reason for thinking that humanity might be in danger. Briefly, SP could be described as: 'Superintelligent machines are feared to be too dumb to possess commonsense.'"

In other words, why would it be that an AI that we have deemed as achieving super-intelligence be so dimwitted that it would fall into these simpleton mental traps? As such, the doomsday scenarios ought to be eyed with a grain of salt, since they at times make assumptions in favor of what a super-intelligence might do, while simultaneously treating the super-intelligence as sub-par intelligence in what it might do.

In this section, *The Singularity* has been briefly expounded to showcase that it is a concept that has been in existence for a considerable while (at least seventy years or so), it is a topic of fluidity and multiple definitions, and that it posits quite serious and significant aspects about the future of AI and the future of humanity. We do not know that it will happen, and we do not know that it will not happen, and yet it is certainly worthwhile contemplating as it bodes for substantive impacts on society if it does indeed occur. Though some technologists are at times focused solely on the challenges and enthralling feat of developing AI to the point of *The Singularity*, there is a great deal of handwringing and concern among those of the AI community about the matter. This is noteworthy since there is often criticism of technologists that they fail to consider the Frankenstein-like potential outcomes of their work [36] [39], which does assuredly happen, yet the special case of *The Singularity* seems to have brought forth an awareness that pushing AI to such an extent requires consideration on what the results might portend.

## 2 Legal Singularity

In this section, the conceptual underpinnings of a potential Legal Singularity are explored. This is undertaken by first examining what the Legal Singularity might consist of, and then identifying how the Legal Singularity leverages *The Singularity* and also what is either omitted or being added beyond the conventional scope of *The Singularity*.

### 2.1 Defining Legal Singularity

The research by Alarie [1] provides a cornerstone indication of what a Legal Singularity might constitute. In brief, he asserts that the expansion of today's Machine Learning capabilities entailing predictive and pattern matching facilities will grow stronger and be fed by masses of data about the law, doing so in an increasingly recursive fueled manner [1]: "The first is that technological progress continues to generate more data. The second is that our methods for analyzing data continue to improve due to increases in computing power and better methods of machine learning."

This would seem to be a phenomenon that would gradually and inexorably evolve and emerge, rather than any kind of overnight or instantaneous type of intelligence explosion. Furthermore, there is nothing overtly indicative that the resulting AI would be of a super-intelligence capacity. It would seem to be computationally impressive and extensive, though not somehow extending beyond the scope of everyday human intelligence as we understand such intelligence to be. Indeed, Alarie makes explicit that *The*

*Singularity* is a provocateur that led to the conceiving of a Legal Singularity, yet does not necessarily embrace the various keystones thereof: “The legal singularity is inspired by and different from the idea of the technological singularity popularized by the futurist Ray Kurzweil. The technological singularity refers to the stage when machines themselves become capable of building ever more capable and powerful machines, to the point of an intelligence explosion that exceeds human understanding or capacity to control (technological singularity is akin, then, to superintelligence).”

Ultimately, according to Alarie [1], the Legal Singularity will be achieved or arrived at and impacts to the law and the practice of law will be overwhelmingly demonstrative, once legal certainty is attained: “The legal singularity will arrive when the accumulation of massively more data and dramatically improved methods of inference make legal uncertainty obsolete. The legal singularity contemplates complete law.”

This is a crucial demarcation about the nature of an envisioned Legal Singularity. There is a presumed and explicitly stated arrival point at which the Legal Singularity can apparently be said to have been attained, namely once legal certainty is achieved, or on the other side of the coin, once all legal uncertainty is eliminated. An interesting and quite worth noting aspect of this as a measuring stick is that it perhaps can be utilized to escape the boundaries of necessarily assuming that Machine Learning and the accumulation of data are the required source for the Legal Singularity to be reached. In other words, if the Legal Singularity is principally defined as the attainment of pure legal certainty, we might then set aside how we got there, and be willing to consider other means by which that target of legal certainty could be attained. There is no need to cling per se to or be anchored to the assumption that it might be due to Machine Learning and the accumulation of data, and there might other explanations that give rise to the Legal Singularity (though the explanation of utilizing Machine Learning and the vast collection of data seems most convincing, today, given what we know about AI as of today).

In an overall sense, the Legal Singularity is defined as an outcome. The outcome is the state at which the law is entirely certain and there is no uncertainty remaining.

Various phrases have arisen to depict this potentiality, including:

- Complete law
- Seamless legal order
- Self-executing legal system
- Completely specified legal system
- Functionally complete law
- Etc.

Another corresponding set of elements underlying this conception of the Legal Singularity is that it will of necessity allow universal access and real-time access to the law, which Alarie explains as: “The legal singularity contemplates the elimination of legal uncertainty and the emergence of a seamless legal order, universally accessible in real-time. In the legal singularity, disputes over the legal significance of agreed facts will be rare. They may be disputes over facts, but the once found, the facts will map on to clear legal consequences. The law will be functionally complete.”

And the Legal Singularity will be in existence within specific domains of the law, along with inevitably occurring in all areas of the law. Alarie uses tax law as an exemplar of a particular domain of law, from which we can generalize across all subdomains of law [1]: “I predict that coming decades will witness three gradual transitions as the legal singularity draws nearer: (1) improved dispute resolution and access to justice in tax law, primarily through the transition from our current reliance on standards (adjudicated *ex post*) to greater reliance on query-able systems of complex rules (knowable *ex ante*); (2) a transition to superior and increasingly more complete specifications of tax law (*i.e.*, a gradual transition from the complex, unwieldy, uncoordinated tax systems of today to tax systems that are massively complex and yet precisely and effectively distribute benefits and burdens); and, (3) with the realization of the legal singularity, a complete specification of tax law (and, indeed, all the other areas of law), which will thenceforth remain (more or less) in positive and normative equilibrium.”

Two additional key facets seem to be given notable consideration. One is that the Legal Singularity will be beneficial to justice and provide a presumably better consequence for society concerning the law [1]: "Ultimately, I believe these developments will result in the "legal singularity" which results in a more or less positively and normatively stable legal system."

Meanwhile, a noted downside to the Legal Singularity will be that it would render the law as less scrutable, perhaps even inscrutable, and thus have correspondingly negative consequences [1]: "The apotheosis of the legal system will be extraordinarily complex and will be beyond the complete understanding of any person." This latter point of a lack of explainability or inability to undertake interpretation is sometimes referred to as a form of computational irreducibility, of which the law would be considered a type of black box in the instantiation of the Legal Singularity.

As a quick recap of the major elements of a postulated Legal Singularity:

- Outcome-based
- Posited on achieving absolute legal certainty
- Will occur gradually, subdomains at a time
- Will ultimately occur across all of law
- Leads to a more stable legal approach
- Law becomes "black box" inscrutable
- Arrival likely occurs via advances in AI and data
- Does not seem to require an intelligence explosion
- Does not seem to require super-intelligence
- Loosely inspired by *The Singularity*

Other researchers have sought to identify both strengths and weaknesses in the case being made for a Legal Singularity. Some view with significant doubt that a Legal Singularity would necessarily produce the anticipated benefits and might instead have substantive adverse consequences. Weber [63] for example postulates that there might be (at least) a twofold threat emerging from a Legal Singularity: (1) it would institutionalize existent biases of the legal system, (2) it would treat human rights as though people are merely atomized data points. Both of those adverse consequences could turn the populace away from embracing a Legal Singularity amid severe qualms about the overturning of the existent normative.

These points are worthwhile to further explore. Per Weber [63], he suggests that we envision an AI system in the future, called Singulatim, and for which it embodies the Legal Singularity. Consider what might happen with Singulatim. There is already today an increasing awareness that contemporary Machine Learning algorithms tend to carryover biases that are inherently in the data used to train the AI systems. If a dataset has been collected that was based on prior decisions containing racist biases, and the Machine Learning patterns to that data, the result is likely to be an AI system that then incorporates and utilizes those biases. There is no common-sense or reasoning by the AI about what the data contains. Furthermore, the AI might have mathematically patterned to the data in an obscure and complex manner, making it nearly impossible to ferret out whether biases are now within the AI system. As stated by Weber: "The first [threat] is essentially critical: that the Singulatim software, in learning from how the legal system works, would institutionalize algorithmically the existing inequalities in the way the legal system treats its subjects."

In the matter of the second major threat of a Legal Singularity, Weber [63] emphasizes that since the AI does not have any cognition or human intelligence per se, and it is merely a Machine Learning algorithm that computationally is doing pattern matching, the result is that humans being subject to the laws are being reduced from being considered as sentient beings and instead be treated as data points in a computational machine [63] "The second threat, on the other hand, strikes even deeper at the rule of law. The problem here is not that the Singulatim software cements in place some extra-legal hierarchy; instead, the issue is that the basic terms of universal rights might cease to make sense in the face of an epistemological shift that allows the law to only see atomized data points where it used to see integral, individual legal subjects." This raises primary and legal core questions that can be formulated by what Weber [63] describes as strong-form theorists and weak-form theorists: "Simplifying only somewhat, strong-form theorists pose the question *Does the legal system protect against arbitrary government power and thereby promote liberty?*, while weak-form theorists ask *Does the legal system promote and maintain social order?"*

In essence, perhaps ironically, the Legal Singularity might usurp and gut or undermine the essence of our approach to law and justice: "In those circumstances, the formerly latent tension holding together the rule of law and the universality principle would explode, destroying any normative force the latter was thought to impart to the former. To adopt the terminology in which legal futurism is often celebrated, the tension, no longer suppressed, would then *disrupt* the very foundations of the rule of law."

These weaknesses underlying the Legal Singularity are subtly predicated on a key facet that otherwise might have seemed not quite so consequential in this discussion so far, namely that the Legal Singularity is portrayed as being shaped by the Machine Learning that we conceive of today. If we reinvigorate the Legal Singularity by bringing from *The Singularity* that there might be a richer embodiment of human intelligence in AI, possibly even a super-intelligence, presumably the twofold threat is no longer quite as emboldened. Here's why. Suppose that the Singulatim had the capacity of human intelligence and therefore might be able to detect and overcome the biases inherent in the underlying data of the law. Furthermore, the Singulatim in the case of the second threat would not necessarily be configuring the status of humans as data points per se, at least to the degree that if human intelligence was equivalence in the AI that there would be some representation beyond the simple data basis that computers today might be ascribed. In essence, the argument made about the Legal Singularity as being limited to ingratiating biases and treating people as mere numbers is predicated on the assumption that the Legal Singularity will be composed of Machine Learning as we make use of it today. By shifting toward a broader futuristic perspective, and an AI of a much greater capability, such an argument does not necessarily continue to hold (that's not to mean that the AI would assuredly obviate biases and nor that it would not treat people as atomized data, only that it opens the door to the possibility that those key facets might not necessarily occur).

As earlier mentioned, there is the Singularity Paradox, proffering the conundrum that some futurists at times are willing to ascribe to AI that it will be something of human intelligence or super-intelligence, and yet in the same context will portray or assume that the AI is to be dimwitted or fail to exhibit intelligence. In the conception of the Legal Singularity, by seemingly removing the capacity of reaching human intelligence (let alone super-intelligence), a Pandora's box of concerns is readily opened widely. It might seem that an easy solution would be to reinsert into the Legal Singularity that AI of human intelligence levels will be required, in which case the "easy and obvious" problems inherent in the Legal Singularity could be explained as overcome by the intrinsic intelligence of the AI involved. This does not suggest that other problems will not ergo arise, only that the ones associated with any automation that is less than the level of human intelligence would typically contain.

Skeptics though tend to deride the requirement of achieving human intelligence (or better) in such matters. First, whether AI can achieve human intelligence (or better) remains an unanswered question and might never be achieved. Thus, if the Legal Singularity did rely upon the assumption that human intelligence (or better) was a necessity for Legal Singularity, it would put the Legal Singularity into the same murky waters as *The Singularity* as to whether this will arise and if ever so. Second, some have a distaste for employing what they consider a magical fix, as it were, consisting of the assumption that human intelligence in AI will be achieved (this is seen as a "insert miracle here" kind of predicate). The viewpoint is that any kind of future can be devised by simply relying upon an amorphous and yet to be proven achievement of human intelligence into a machine.

As an example of this kind of conundrum, it is relatively straightforward to pick apart the Legal Singularity in terms of its potential impacts by aiming at the already known limitations and shortcomings of today's automation including Machine Learning. In the research by Deakin and Markou [18] they point out that law is a social institution and the Legal Singularity would operate in a social vacuum since it is computationally based (as we know of it today): "But if mathematical logic cannot capture the 'situation-specific understanding' of legal reasoning and the complexity of the social world it exists in—at least to any extent congruent with how natural language categories cognize social referents and character of meaning—the hypothetical totalization of 'AI judges' implied by the legal singularity would instantiate a particular view of law: one in which legal judgments are essentially deterministic or probabilistic outputs, produced on the basis of static or unambiguous legal rules, in a societal vacuum. This would deny, or see as irrelevant, competing conceptions of law, in particular the idea that law is a social institution, involving socially constructed activities, relationships, and norms not easily translated into numerical functions. It would also turn a blind eye to the reality that legal decision making involves an exercise of power which is both material and, in Pierre Bourdieu's sense, 'symbolic '."

Overall, the Legal Singularity as conceptualized without some apparent semblance of human intelligence (or better) in the underlying AI is endlessly vulnerable to any number of shotgun or scattergun attacks as to being unable to rise above the limitations already known (and some likely yet to be surfaced) about contemporary versions of AI. This is not an attempt to have Legal Singularity switchover to embracing a more powerful semblance or sense of AI, and only pointing out the quagmire associated with a definition of Legal Singularity that resides dependent upon and impotent due to the assumption of today's AI mechanizations. To be fair, the original concept does provide a type of escape clause by emphasizing that the Machine Learning would be based on better methods than used today, which is a crucial point that seems to be at times lost or lessened in criticisms of the conception of Legal Singularity. In any case, as will be discussed in Section 4, one means to cope with this difficulty is to recast the Legal Singularity in terms of the autonomous levels of AI Legal Reasoning, providing a path toward conceiving of the Legal Singularity across a spectrum of what AI might become.

## 3 Autonomous Levels of AI Legal Reasoning

In this section, a framework for the autonomous levels of AI Legal Reasoning is summarized and is based on the research described in detail in Eliot [25].

These autonomous levels will be portrayed in a grid that aligns with key elements of autonomy and as matched to AI Legal Reasoning. Providing this context will be useful to the later sections of this paper and will be utilized accordingly.

The autonomous levels of AI Legal Reasoning are as follows:

Level 0: No Automation for AI Legal Reasoning
Level 1: Simple Assistance Automation for AI Legal Reasoning
Level 2: Advanced Assistance Automation for AI Legal Reasoning
Level 3: Semi-Autonomous Automation for AI Legal Reasoning
Level 4: Domain Autonomous for AI Legal Reasoning
Level 5: Fully Autonomous for AI Legal Reasoning
Level 6: Superhuman Autonomous for AI Legal Reasoning

See **Figure A-1** for an overview chart showcasing the autonomous levels of AI Legal Reasoning as via columns denoting each of the respective levels.

See **Figure A-2** for an overview chart similar to Figure A-1 which alternatively is indicative of the autonomous levels of AI Legal Reasoning via the rows as depicting the respective levels (this is simply a reformatting of Figure A-1, doing so to aid in illuminating this variant perspective, but does not introduce any new facets or alterations from the contents as already shown in Figure A-1).

### 3.1.1 Level 0: No Automation for AI Legal Reasoning

Level 0 is considered the no automation level. Legal reasoning is carried out via manual methods and principally occurs via paper-based methods.

This level is allowed some leeway in that the use of say a simple handheld calculator or perhaps the use of a fax machine could be allowed or included within this Level 0, though strictly speaking it could be said that any form whatsoever of automation is to be excluded from this level.

### 3.1.2 Level 1: Simple Assistance Automation for AI Legal Reasoning

Level 1 consists of simple assistance automation for AI legal reasoning.

Examples of this category encompassing simple automation would include the use of everyday computer-based word processing, the use of everyday computer-based spreadsheets, the access to online legal documents that are stored and retrieved electronically, and so on.

By-and-large, today's use of computers for legal activities is predominantly within Level 1. It is assumed and expected that over time, the pervasiveness of automation will continue to deepen and widen, and eventually lead to legal activities being supported and within Level 2, rather than Level 1.

### 3.1.3 Level 2: Advanced Assistance Automation for AI Legal Reasoning

Level 2 consists of advanced assistance automation for AI legal reasoning.

Examples of this notion encompassing advanced automation would include the use of query-style Natural Language Processing (NLP), Machine Learning (ML) for case predictions, and so on.

Gradually, over time, it is expected that computer-based systems for legal activities will increasingly make use of advanced automation. Law industry technology that was once at a Level 1 will likely be refined, upgraded, or expanded to include advanced capabilities, and thus be reclassified into Level 2.

### 3.1.4 Level 3: Semi-Autonomous Automation for AI Legal Reasoning

Level 3 consists of semi-autonomous automation for AI legal reasoning.

Examples of this notion encompassing semi-autonomous automation would include the use of Knowledge-Based Systems (KBS) for legal reasoning, the use of Machine Learning and Deep Learning (ML/DL) for legal reasoning, and so on.

Today, such automation tends to exist in research efforts or prototypes and pilot systems, along with some commercial legal technology that has been infusing these capabilities too.

### 3.1.5 Level 4: Domain Autonomous for AI Legal Reasoning

Level 4 consists of domain autonomous computer-based systems for AI legal reasoning.

This level reuses the conceptual notion of Operational Design Domains (ODDs) as utilized in the autonomous vehicles and self-driving cars levels of autonomy, though in this use case it is being applied to the legal domain [17] [18] [20].

Essentially, this entails any AI legal reasoning capacities that can operate autonomously, entirely so, but that is only able to do so in some limited or constrained legal domain.

### 3.1.6 Level 5: Fully Autonomous for AI Legal Reasoning

Level 5 consists of fully autonomous computer-based systems for AI legal reasoning.

In a sense, Level 5 is the superset of Level 4 in terms of encompassing all possible domains as per however so defined ultimately for Level 4. The only constraint, as it were, consists of the facet that the Level 4 and Level 5 are concerning human intelligence and the capacities thereof. This is an important emphasis due to attempting to distinguish Level 5 from Level 6 (as will be discussed in the next subsection)

It is conceivable that someday there might be a fully autonomous AI legal reasoning capability, one that encompasses all of the law in all foreseeable ways, though this is quite a tall order and remains quite aspirational without a clear cut path of how this might one day be achieved. Nonetheless, it seems to be within the extended realm of possibilities, which is worthwhile to mention in relative terms to Level 6.

### 3.1.7 Level 6: Superhuman Autonomous for AI Legal Reasoning

Level 6 consists of superhuman autonomous computer-based systems for AI legal reasoning.

In a sense, Level 6 is the entirety of Level 5 and adds something beyond that in a manner that is currently ill-defined and perhaps (some would argue) as yet unknowable. The notion is that AI might ultimately exceed human intelligence, rising to become superhuman, and if so, we do not yet have any viable indication of what that superhuman intelligence consists of and nor what kind of thinking it would somehow be able to undertake.

Whether a Level 6 is ever attainable is reliant upon whether superhuman AI is ever attainable, and thus, at this time, this stands as a placeholder for that which might never occur. In any case, having such a placeholder provides a semblance of completeness, doing so without necessarily legitimatizing that superhuman AI is going to be achieved or not. No such claim or dispute is undertaken within this framework.

## 4 Legal Singularity Multidimensionality, Alignment with LoA AILR

This section combines the prior sections respective discussions about or pertaining to the Legal Singularity, doing so in two key ways: (a) Showcase how the Legal Singularity aligns with the autonomous levels of AI Legal Reasoning, and (b) Indicate the multidimensionality of the Legal Singularity by conducting a parametric analysis. This then provides the core contributions of this paper, and Section 5 then offers concluding remarks and recommendations for further research on these matters.

### 4.1 Legal Singularity and the LoA AILR

Existing indications about the Legal Singularity seem to imply that the Legal Singularity will potentially arise at a particular point in time. Another viewpoint would be to consider that the Legal Singularity will arise in a series of stages or phases.

As shown in **Figure B-1**, the autonomous levels for AI Legal Reasoning are presented, and included in the lower region of the chart is a projected depiction of the Legal Singularity. This overlay provides a convenient means of portraying the possibility that the Legal Singularity will gradually emerge and evolve over time.

At the levels less than 3, there is no expectation of the Legal Singularity and thus it is a grayed-out indication. Even though an argument could be made that during Level 1 and Level 2 there is some amount of groundwork being laid for the seeding and later growth of the Legal Singularity, this seems to be a pre-seeding effort and not especially noteworthy for highlighting for the overall anticipated inception of the Legal Singularity.

During Level 3, the Legal Singularity begins to take some palpable shape, doing so during the advanced AI efforts that consist primarily of prototypes and research-based activities of applying AI to the law. This tryout status will aid in ascertaining the viability of a Legal Singularity and likely to reveal the feasibility of a Legal Singularity occurring all told.

Assuming that the Legal Singularity is viable, there is a denoted Stage A that might occur during Level 4, and a Stage B that might occur during Level 5, and a Stage C that might occur during Level 6. This indication should not be interpreted as a signal or prophecy that the Legal Singularity will indeed happen, since that is not the purpose or intent of this chart, and instead the viewpoint is that if the Legal Singularity were to arise that it might do so in the staged manner presented in the chart.

At Level 4, the Legal Singularity would be taking hold at various subdomains of the law, such as a Legal Singularity in real estate law, family law, and so on.

At Level 5, the Legal Singularity would be across all subdomains of law and therefore encompass all of the law.

At Level 6, the Legal Singularity would be akin to Level 5 in that it would encompass all of the law and have an added aspect that the AI would be superhuman or consist of super-intelligence. As already noted earlier in this paper, it is unknown as to what the superhuman or super-intelligence might consist of, and thus highly speculative to assert what this might achieve in the case of Legal Singularity. In any case, since some believe a superhuman capacity might someday exist in AI, the Level 6 accounts for that possibility and similarly, the Legal Singularity accounts for the possibility too via an indicated Stage C.

Overall, the Legal Singularity is aligned with the levels of autonomy (LoA) of the AI Legal Reasoning (AILR) in this manner:

- **Level 0: <*not noteworthy*>**
- **Level 1: <*not noteworthy*> (pre-seed)**
- **Level 2: <*not noteworthy*> (pre-seed)**
- **Level 3: Tryout (pre-stage)**
- **Level 4: Stage A**
- **Level 5: Stage B**
- **Level 6: Stage C**

### 4.2 Multidimensionality of Legal Singularity

Law research and the scholarly literature about the Legal Singularity have tended to *silently* encompass various dimensions underlying the Legal Singularity, meaning that those research efforts have not usually explicitly stated the dimensions being considered. It is contended here that those dimensions are in fact overtly identifiable and distinct, and of necessity should be explicitly stated. In essence, the dimensions have often been treated implicitly, serving as hidden assumptions, and not directly and purposely addressed per se.

This lack of overtly naming the dimensions can confound discussions about the Legal Singularity. For example, research examining the Legal Singularity might fail to name a dimension and make essential unstated assumptions about its nature impacting the Legal Singularity. This omission or hidden assumption renders the research less informative and can undermine progress on explicating the Legal Singularity more fully. Furthermore, trying to compare one research effort on Legal Singularity to another can become onerous and unnecessarily argumentative due to a lack of stated dimensions, including the underlying assumptions the research authors are each respectively making regarding each such dimension.

In reviewing the prior research on Legal Singularity, a dozen key dimensions have been identified. These are not all of the dimensions that might be conceived of, and merely a considered core set, though nonetheless provides a helpful starter and foundational means to further explore the multidimensionality.

**Figure B-2** indicates the dozen identified dimensions. Those identified dimensions consist of:

- **Alignment of Legal Singularity**
- **Pace of Legal Singularity**
- **Capability of Legal Singularity**
- **Cornerstone of Legal Singularity**
- **Scope of Legal Singularity**
- **Legal Profession and Legal Singularity**
- **Social Outcome of Legal Singularity**
- **Justice and Legal Singularity**
- **Paradoxes of Legal Singularity**
- **Defeaters of Legal Singularity**
- **Explainability of Legal Singularity**
- **Control of Legal Singularity**

Each of these dimensions will be discussed in the next subsections.

Note that the dimensions are not numbered, which is done purposely, since there is some apprehension that if they were shown in a numbered list it might imply a sense of priority or ranking among the dimensions. It is intended that the dimensions are to be considered without any overall ranking or priority and that they are all equal as elements or parameters of the Legal Singularity. That being said, there is certainly the usefulness of considering whether some dimensions are "more equal than others" and could be considered having greater weight in the emergence of the Legal Singularity or perhaps when assessing the potential impacts of the Legal Singularity. Thus, additional research could indeed opt to proffer weights or rankings to the dimensions, but doing so in this paper would seem to possibly undermine the crucial premise and distract from the overarching concept that there are dimensions and that those dimensions are worthy of attention (no need to distract from that premise by also simultaneously trying to tackle a ranking dispute too).

**Figure B-3** indicates the dimensions as shown in a range measurement chart.

Each of the dimensions can be assigned a measuring element, doing so to further amplify and make visible the assumptions underlying the utilization of the dimension when discussing the Legal Singularity as a concept and potential phenomena. In this chart of Figure B-3, a range portrayal is used, indicating some semblance of varying assumptions about the dimension. Do not misinterpret the chart by assuming that the ranges are somehow all equal or comparable, which they are most decidedly not. The ranges are dimension specific. Furthermore, the ranges and the dimensions are not shown in any particular order that would imply prioritization or ranking (as earlier so pointed out).

The intent of the Figure B-3 chart will become more evident when used as a means of comparing how different research on Legal Singularity has tended to characterize the Legal Singularity and can aid too in making explicit the implicit assumptions of those research efforts. This will be further discussed in the subsections of this section. Note too that there is nothing magical or dogmatic about the range indicators, such that additional research is likely to indicate other means of specifying the ranges and their utility in being measured and compared.

**4.2.1 Alignment of Legal Singularity**

Alignment of Legal Singularity generally refers to its timing with respect to *The Singularity*.

Some assert that the Legal Singularity will occur and perhaps can only occur upon or after *The Singularity* has happened, whilst others claim that the Legal Singularity can occur before *The Singularity* and indeed there is not any necessity that The Singularity needs to ever happen (concerning the occurrence of the Legal Singularity).

**4.2.2 Pace of Legal Singularity**

Pace of Legal Singularity generally refers to the speed at which the Legal Singularity will emerge or arise.

Some assert that the Legal Singularity will playout gradually, step by step, over an elongated time, while others indicate that as per many beliefs about *The Singularity* that there will be a sparked moment or instantaneous emergence rather than a gradual one.

**4.2.3 Capability of Legal Singularity**

Capability of Legal Singularity generally refers to the magnitude of intelligence requisite for the onset of the Legal Singularity.
Some assert that the Legal Singularity will be enabled by AI and Machine Learning that is either at the human level of intelligence or akin to human intelligence but perhaps less so in certain respects, while others believe that the Legal Singularity will require a superhuman or super-intelligence capacity by the AI.

**4.2.4 Cornerstone of Legal Singularity**

Cornerstone of Legal Singularity generally refers to the crucial component of certainty, considered a cornerstone upon which Legal Singularity is founded.

Some assert that Legal Singularity will only be considered as emerged when legal certainty is achieved as an absolute, thus presumably eliminating all legal uncertainty, while others believe that some amount of legal uncertainty can remain and yet nonetheless still have the achievement of Legal Singularity.

**4.2.5 Scope of Legal Singularity**

Scope of Legal Singularity generally refers to the aspect of how much of the law will be encompassed by the Legal Singularity.

Some assert that the Legal Singularity will entail all of the law, while others indicate it could be instead selected subdomains of the law, for which both viewpoints might be in agreement if it is said that this will evolve, though these views could be in disagreement if it is stated as a winner-take-all that the Legal Singularity only arises when all of the law has been included.

**4.2.6 Legal Profession and Legal Singularity**

Legal Profession and Legal Singularity generally refers to the notion that the Legal Singularity might dramatically impact the legal profession in terms of the need for and employment of human legal professionals.

Some assert that a Legal Singularity might be seen as an augmentation to the legal profession, thus to some degree still employing and requiring the use of human legal professionals, whilst others suggest that the legal profession might be “wiped out” entirely and be replaced by AI as part of a Legal Singularity emergence.

**4.2.7 Social Outcome of Legal Singularity**

Social Outcome of Legal Singularity generally refers to the societal result of a Legal Singularity.

Some assert that a Legal Singularity might cause the law to become a societally oppressive tool and produce a Dystopian future, whilst others believe that the Legal Singularity will provide a societally uplifting capacity that will lead to a Utopian style future.

### 4.2.8 Justice and Legal Singularity

Justice and Legal Singularity generally refers to the impacts that the Legal Singularity would have on the principles of justice, equity, and fairness.
Some assert that the Legal Singularity might lessen justice, reduce equity, and usurp fairness, while others argue that it could instead boost justice, increase equity, and provide greater assurance of fairness.

### 4.2.9 Paradoxes of Legal Singularity

Paradoxes of Legal Singularity generally refer to the same notion as Singularity Paradoxes of *The Singularity* (as explicated in the prior sections) but as applicable to the Legal Singularity.

Some assert that if the Legal Singularity can eliminate legal uncertainty that it ergo is implausible to attack Legal Singularity for other potential failings since it would presumably be as strong in those other respects, while others argue that there are potential weak spots nonetheless and other problematic aspects that are detectable and decidedly not paradoxical.

### 4.2.10 Defeaters of Legal Singularity

Defeaters of Legal Singularity generally refer to the same notion as Defeaters with respect to *The Singularity* (as explicated in prior sections) but as applicable to the Legal Singularity.

An overarching question often posed about *The Singularity* it is inevitable or whether mankind will explicitly ascertain whether it will happen; likewise, the Legal Singularity can be said to subject to the same conditions, namely that there might be a plethora of aspects that could either delay the Legal Singularity or render it never to arise, and for which might be led by those within the law industry or those outside of the legal profession.

### 4.2.11 Explainability of Legal Singularity

Explainability of Legal Singularity generally refers to the aspect of whether the law and the legal mechanizations thereof will be explainable in the emergence of the Legal Singularity.

Some assert that the Legal Singularity will end-up rendering the law as inscrutable, whilst others contend that the law might become more visible, more explainable, and better understood as a result of the Legal Singularity.

### 4.2.12 Control of Legal Singularity

Control of Legal Singularity generally refers to the amount of control of the Legal Singularity by mankind versus automation.

Some assert that the Legal Singularity could produce an automation-based form of legal justice that becomes detached from humanity and might end-up with essentially AI being in control, whilst others argue that the touch of mankind would remain firmly on the wheels of justice and be overseeing and able to fully control the legal automation or autonomy.

## 4.3 Examples of the Legal Singularity Dimensions Chart

To illuminate the utility of having the Legal Singularity dimensions explicitly arrayed, consider how the dimensionality chart can be productively utilized.

**Figure B-4** shows an example of the Legal Singularity dimensional chart as marked for a scenario labeled simply as Example 1A.

Envision that a researcher has examined the Legal Singularity and offered various nuances and arguments in favor of or opposition to other prior research.

Likely, there are numerous base assumptions that the researcher has made about the Legal Singularity.

By using the Legal Singularity dimensional chart, we can make explicit the assumptions being made. As shown in Figure B-4, the research is essentially postulating that:

- Alignment of Legal Singularity: **Before AI Singularity**
- Pace of Legal Singularity: **Gradual**
- Capability of Legal Singularity: **Human Intelligence (minimal needed)**
- Cornerstone of Legal Singularity: **Absolute Legal Certainty**
- Scope of Legal Singularity: **Subdomains (leading toward all)**
- Legal Profession and Legal Singularity: **Augmentation**
- Social Outcome of Legal Singularity: **Dystopian and Utopian (mixed)**
- Justice and Legal Singularity: **More Equity & Fairness (tends toward)**
- Paradoxes of Legal Singularity: **Some**
- Defeaters of Legal Singularity: **Within Law**
- Explainability of Legal Singularity: **Inscrutable**
- Control of Legal Singularity: **By Mankind**

Some notable facets to keep in mind are that the ranges are not intended to be numerical, which some might desire, such as numbering each of the markers between the ranges. It is not intended that the chart would be used in such a fashion, and once again if done as such might distract from its overall utility. Likewise, it is essentially inappropriate to try and state that a range endpoint is a descriptor when the diamond marker is somewhere along the given spectrum. In that sense, even the above indicates that the pace of legal singularity is "Gradual" provides a somewhat misleading and flat indication of what the actual marking consisted of, which was primarily toward gradual but with some semblance of leaning slightly toward the instantaneous.

**Figure B-5** is another example, labeled as Example 1B.

This example showcases a circumstance whereby the research being analyzed for its base assumption across each of the dimensions has tended toward the extremes of the ranges. If there was interest in comparing the research depicted by Example 1A with Example 1B, it would be relatively straightforward to then compare the two as based on the assumptions they each respectively are making about how the Legal Singularity is to be considered.

## 5.0 Additional Considerations and Future Research

There is a myriad of additional considerations that arise from this discussion about Legal Singularity and further research is amply warranted.

As an example of open topics, consider the aspect that Legal Singularity appears to be predicated principally on the singular dimension or parameter entailing legal certainty (or if so preferred, legal uncertainty). This focus on an individual dimension as the particular underpinning can be viewed as problematic for a variety of demonstrative reasons, as I will outline next.

Concerns about the certainty dimension include these facets:

- **Oversized Requirement for Purity of Certainty**. A purity assumption of attaining absolute legal certainty as a precondition for Legal Singularity is potentially an insurmountable hurdle since it presumably precludes any amount of legal uncertainty, even the most infinitesimal trace, and this seems a prohibitive directive that does not allow for the likely wavering or fluctuation of and between states of legal certainty and legal uncertainty. As such, apparently, as long as there is any semblance of legal uncertainty, Legal Singularity cannot be deemed as having been reached and nor maintained, and the question arises whether the complete expungement of legal uncertainty shall be feasible.

- **Assumption of Exclusively Deterministic Algorithms**. The manner of Machine Learning and AI that will produce the legal certainty seems to be based on a form of deterministic algorithms, exclusively, as though there is no inclusion of non-deterministic algorithms. There does not seem to be any corroborated basis in the defining of Legal Singularity to support such a claim or assertion of this presumed deterministic nature. As such, given that non-determinism is a seemingly strong potential in the case of AI and the law, and perhaps even a necessary ingredient, this realization then introduces probabilistic behavior, which in turn substantively undermines the tenet of requiring complete and steadfast legal certainty.

- **Progression Toward Legal Uncertainty Rather Than Legal Certainty.** Research by D'Amato [68] asserts that legal certainty is decreasing over time, thus legal uncertainty is rising, and that there is a fallacy among many legal scholars that falsely propagate a myth of legal certainty increasing over time. He suggests that the rules and principles of law tend to splinter and be generative over time and thus engenders legal uncertainty [68]: "Legal certainty decreases over time. Rules and principles of law become more and more uncertain in content and in application because legal systems are biased in favor of unraveling those rules and principles." In his view, legal rules suffer from several maladies [68]: "When I argue that rules unravel over time, I mean that, using any of these extended definitions of the term, a 'rule' becomes increasingly vague, inapplicable, remote, ambiguous, or exception ridden." The Legal Singularity appears to presuppose that due to the (future) AI capacity of predictiveness, the law will become increasingly certain, but this raises at least two considerations. First, this might be a proverbial cart before the horse in that the AI is assumed as somehow leading to certainty and yet the law itself might be inexorably moving intrinsically toward uncertainty. Second, if one assumes that the advent of AI is going to reverse the tendency of the law going toward uncertainty, this needs some robust rationalizing as to why this would of necessity be the case (i.e., it might provide some impetus, but the argument seems to be made is that it will magnetically do so to the degree of achieving ultimate and complete legal certainty).

- **Doctrinaire Belief That Legal Certainty Is The Pinnacle.** On the surface, there is a comforting sense that eliminating all legal uncertainty is a highly desirable outcome and that achieving purity of legal certainty is a proper and crucial goal. But there seems to be more to the tradeoffs between legal certainty and the allowance for some amount of legal uncertainty than otherwise ordinarily meets the eye. As per D'Amato [68], he indicates that though legal certainty is generally and primarily the desirable goal, there is nonetheless still a basis for some value from legal uncertainty: "One may ask, however, whether uncertainty in the law is undesirable. Although I contend that it is, in some cases it might not be." Thus, if the Legal Singularity is the apex, and for which legal certainty underlies it, there would seem to be a need to substantiate how the solidity of legal certainty will overcome those instances for which legal uncertainty is viewed as a positive rather than a negative element.

- **Legal Certainty Is Only One Leg Of The Law Triad:** Focusing solely on legal certainty as the bedrock dimension for Legal Singularity would appear to defy the assertion that legal certainty is part of a triad of the law (which will be elucidated momentarily herein), and thus encompasses only one of three key principles of the law that need to be observed. By many legal scholars, it is generally suggested that the legal triad is akin to a three-legged stool, whereby each leg exists to keep the others in balance, and a stool with but one leg would be unbalanced. Consider this indication of Radbruch's legal precepts as depicted by Leawood [69]: "To complete the concept of law Radbruch uses three general precepts: purposiveness, justice, and legal certainty. Therefore, Radbruch defines law as 'the complex of general precepts for the living-together of human beings' whose ultimate idea is oriented toward justice or equality." In the legal certainty leg of the law, Leawood depicts Radbruch's views in this manner: "Radbruch's final precept is legal certainty. An important part of legal certainty is the justice it provides through, if nothing else, its predictability." This then indubitably supports the importance of legal certainty and bolsters its basis for being at the core of Legal Singularity, but Leawood points out that legal certainty is not an island unto itself: "Certainly, the conflict between legal certainty and justice or between legal certainty and purposiveness is easy to imagine. For example, legal certainty would demand that a law be upheld even though the result would be an unjust application of the law. Therefore, in most cases the content, form, and validity of the law are understood in terms of Radbruch's triad; three equally weighted principles that, while in tension and possibly in contradiction, are found together." In short, how does the envisioned Legal Singularity motivate the triad if the seemingly sole measure is to be based on legal certainty, and as such might lead to serious deficiencies in the other two, namely purposiveness and justice, by overemphasizing and potentially undercutting the tension and dynamics of the triad?

- **Legal Certainty Reliance Upon Legal Rules Versus Legal Principles.** The Legal Singularity would appear to suggest that the advent of AI and Machine Learning will enable encapsulation of legal rules, and in turn, this will lead to the attainment of legal certainty. Essentially, the assumption would appear to be that legal rules will ultimately and unerringly produce legal certainty. Some legal scholars have argued that there are circumstances whereby legal rules can lead to legal certainty, and yet there is also circumstance for which legal principles lead to legal certainty and legal rules do not. Per Braithwaite [67]: "This has been an attempt to develop a theory of legal certainty and to show that questions like whether presumptive positivism is a legal theory that should attract our allegiance depends on testing its empirical claims and assumptions about how rules work. The theory we have come to has three propositions: (1) When the type of action to be regulated is simple, stable and does not involve huge economic interests, rules tend to regulate with greater certainty than principles. (2) When the type of action to be regulated is complex, changing and involves large economic interests: (a) principles tend to regulate with greater certainty than rules; (b) binding principles backing non-binding rules tend to regulate with greater certainty than principles alone; (c) binding principles backing non-binding rules are more certain still if they are embedded in institutions of regulatory conversation that foster shared sensibilities." If the Legal Singularity is foundationally assuming that only legal rules will lead to the desired legal certainty, this would seem to overlook or omit the role of legal principles, but if legal principles are also to be included it raises the corresponding question of how legal certainty is to be attained and legal uncertainty to be eradicated.

These probing questions about the legal certainty dimension are vital to the crux of the Legal Singularity concept. From such questions, it is potentially the case that further refinement and adjustment of the Legal Singularity might be spurred. Additional dimensions might be considered for inclusion or at least for explicit acknowledgment and placement. Asking these kinds of questions is not to be interpreted as a disparaging of the Legal Singularity and instead should be viewed as aiding the future exploration and maturation of the Legal Singularity concept.

This paper has closely examined the postulated Legal Singularity and proffered that such AI and Law cogitations can be enriched by the three facets addressed herein: (1) by dovetailing additionally salient considerations of *The Singularity* into the Legal Singularity, (2) by making use of the in-depth and innovative multidimensional parametric analysis of the Legal Singularity as posited in this paper, and (3) by aligning and unifying the Legal Singularity with the Levels of Autonomy (LoA) associated with AI Legal Reasoning (AILR) as propounded in this paper.

**About the Author**

Dr. Lance Eliot is the Chief AI Scientist at Techbrium Inc. and a Stanford Fellow at Stanford University in the CodeX: Center for Legal Informatics. He previously was a professor at the University of Southern California (USC) where he headed a multi-disciplinary and pioneering AI research lab. Dr. Eliot is globally recognized for his expertise in AI and is the author of highly ranked AI books and columns.

**Figure A-1**

## AI & Law: Levels of Autonomy For AI Legal Reasoning (AILR)

| Level | Descriptor | Examples | Automation | Status |
|---|---|---|---|---|
| 0 | No Automation | Manual, paper-based (no automation) | None | De Facto - In Use |
| 1 | Simple Assistance Automation | Word Processing, XLS, online legal docs, etc. | Legal Assist | Widely In Use |
| 2 | Advanced Assistance Automation | Query-style NLP, ML for case prediction, etc. | Legal Assist | Some In Use |
| 3 | Semi-Autonomous Automation | KBS & ML/DL for legal reasoning & analysis, etc. | Legal Assist | Primarily Prototypes & Research Based |
| 4 | AILR Domain Autonomous | Versed only in a specific legal domain | Legal Advisor (law fluent) | None As Yet |
| 5 | AILR Fully Autonomous | Versatile within and across all legal domains | Legal Advisor (law fluent) | None As Yet |
| 6 | AILR Superhuman Autonomous | Exceeds human-based legal reasoning | Supra Legal Advisor | Indeterminate |

*Figure 1: AI & Law - Autonomous Levels by Rows* *Source Author: Dr. Lance B. Eliot* V1.3

**Figure A-2**

**AI & Law: Levels of Autonomy For AI Legal Reasoning (AILR)**

| | Level 0 | Level 1 | Level 2 | Level 3 | Level 4 | Level 5 | Level 6 |
|---|---|---|---|---|---|---|---|
| **Descriptor** | No Automation | Simple Assistance Automation | Advanced Assistance Automation | Semi-Autonomous Automation | AILR Domain Autonomous | AILR Fully Autonomous | AILR Superhuman Autonomous |
| **Examples** | Manual, paper-based (no automation) | Word Processing, XLS, online legal docs, etc. | Query-style NLP, ML for case prediction, etc. | KBS & ML/DL for legal reasoning & analysis, etc. | Versed only in a specific legal domain | Versatile within and across all legal domains | Exceeds human-based legal reasoning |
| **Automation** | None | Legal Assist | Legal Assist | Legal Assist | Legal Advisor (law fluent) | Legal Advisor (law fluent) | Supra Legal Advisor |
| **Status** | De Facto – In Use | Widely In Use | Some In Use | Primarily Prototypes & Research-based | None As Yet | None As Yet | Indeterminate |

*Figure 2: AI & Law - Autonomous Levels by Columns* *Source Author: Dr. Lance B. Eliot* V1.3

**Figure B-1**

## Legal Singularity: Levels of Autonomy For AI Legal Reasoning (AILR)

| | Level 0 | Level 1 | Level 2 | Level 3 | Level 4 | Level 5 | Level 6 |
|---|---|---|---|---|---|---|---|
| **Descriptor** | No Automation | Simple Assistance Automation | Advanced Assistance Automation | Semi-Autonomous Automation | AILR Domain Autonomous | AILR Fully Autonomous | AILR Superhuman Autonomous |
| **Examples** | Manual, paper-based (no automation) | Word Processing, XLS, online legal docs, etc. | Query-style NLP, ML for case prediction, etc. | KBS & ML/DL for legal reasoning & analysis, etc. | Versed only in a specific legal domain | Versatile within and across all legal domains | Exceeds human-based legal reasoning |
| **Automation** | None | Legal Assist | Legal Assist | Legal Assist | Legal Advisor (law fluent) | Legal Advisor (law fluent) | Supra Legal Advisor |
| **Status** | De Facto – In Use | Widely In Use | Some In Use | Primarily Prototypes & Research-based | None As Yet | None As Yet | Indeterminate |

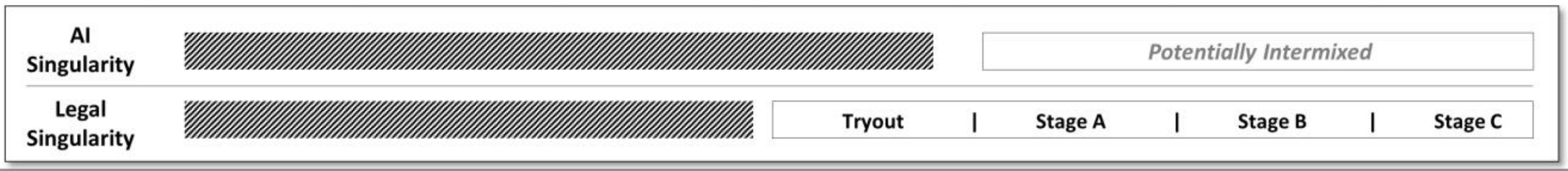

*Figure 1: Legal Singularity - Autonomous Levels of AILR by Columns*    *Source Author: Dr. Lance B. Eliot*

V1.3

**Figure B-2**

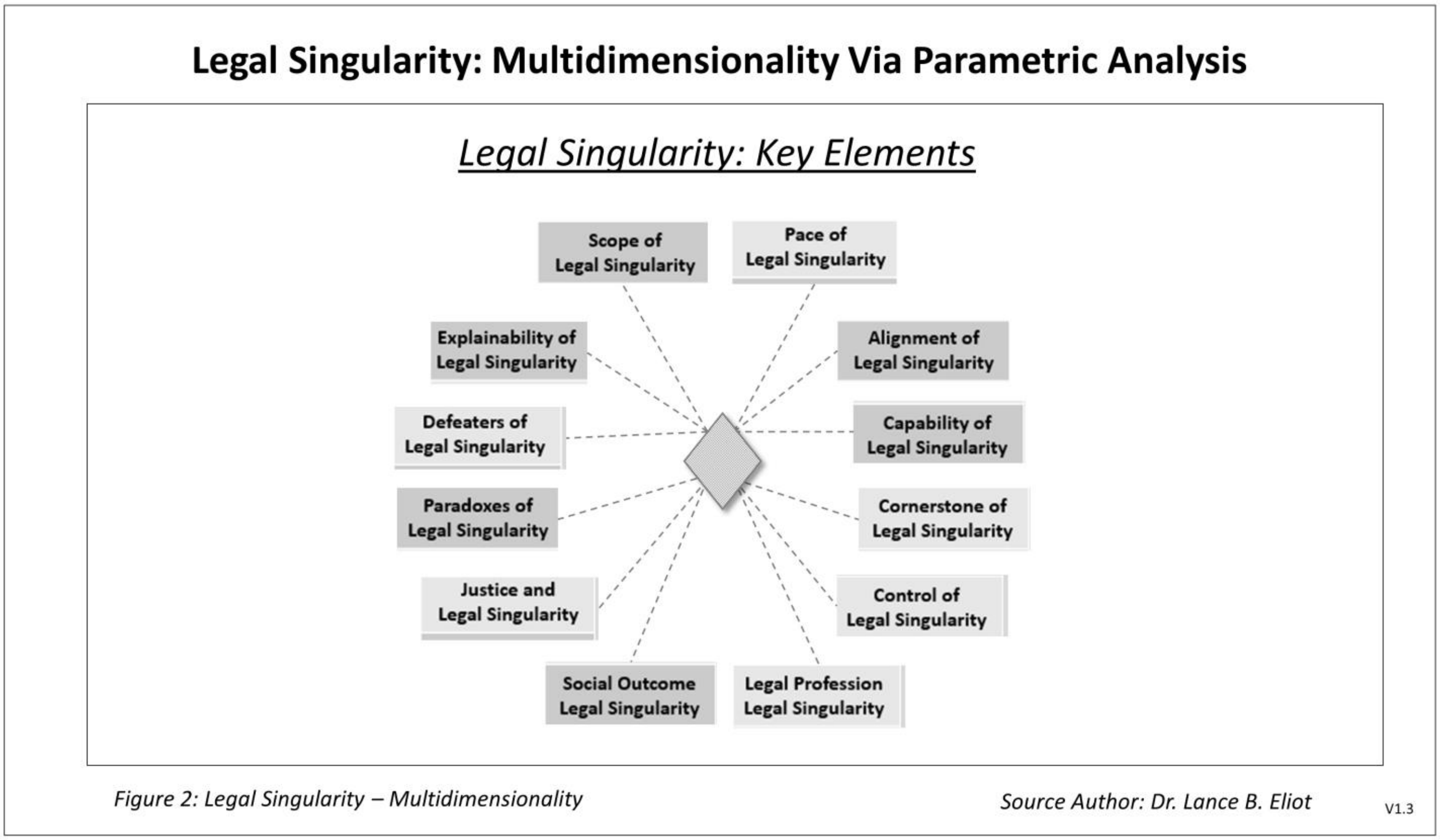
Legal Singularity: Multidimensionality Via Parametric Analysis
Legal Singularity: Key Elements
Scope of Legal Singularity
Pace of Legal Singularity
Explainability of Legal Singularity
Alignment of Legal Singularity
Defeaters of Legal Singularity
Capability of Legal Singularity
Paradoxes of Legal Singularity
Cornerstone of Legal Singularity
Justice and Legal Singularity
Control of Legal Singularity
Social Outcome Legal Singularity
Legal Profession Legal Singularity
Figure 2: Legal Singularity – Multidimensionality
Source Author: Dr. Lance B. Eliot
V1.3

**Figure B-3**

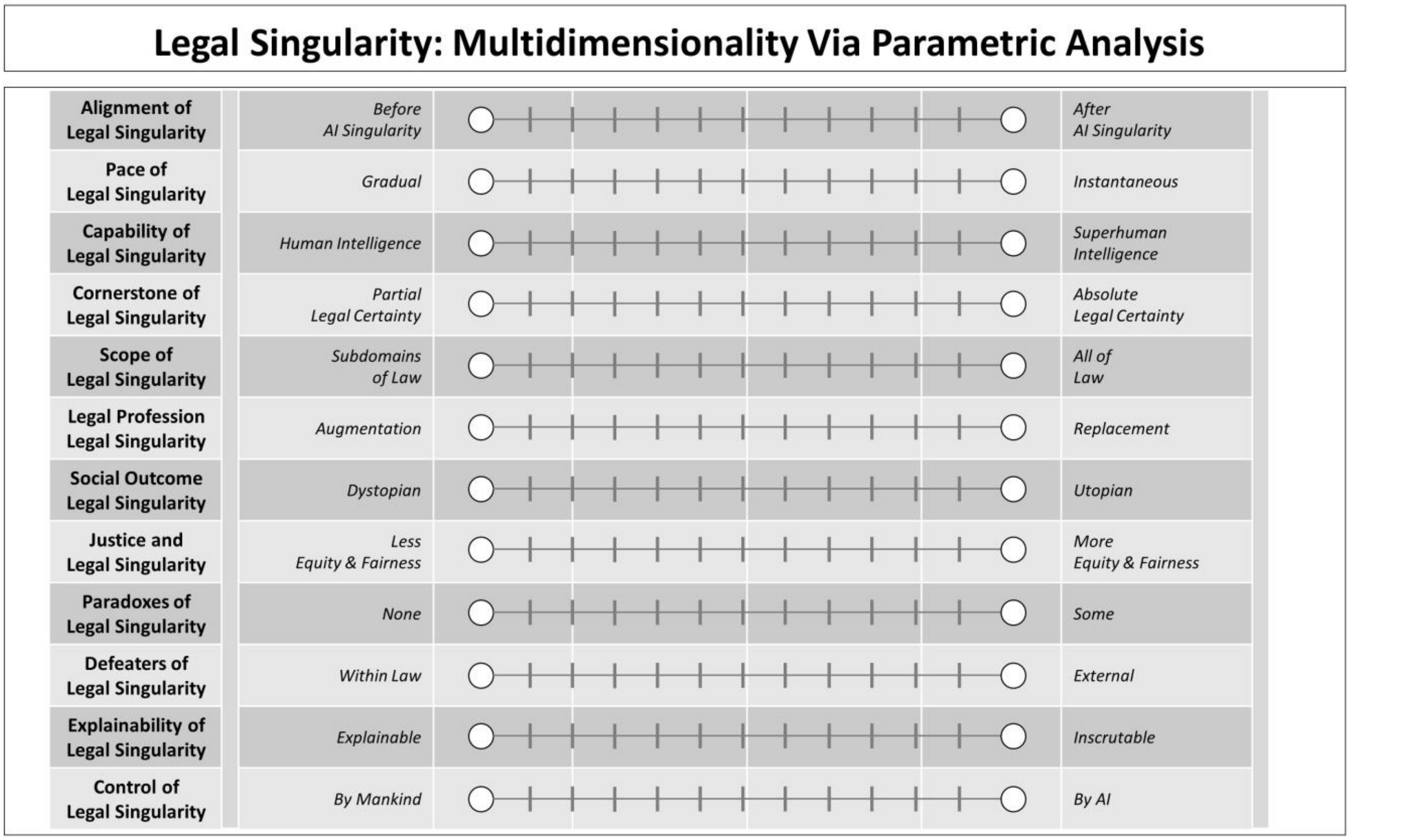

*Figure 3: Legal Singularity Multidimensionality and Parametric Analysis* *Source Author: Dr. Lance B. Eliot*

**Figure B-4**

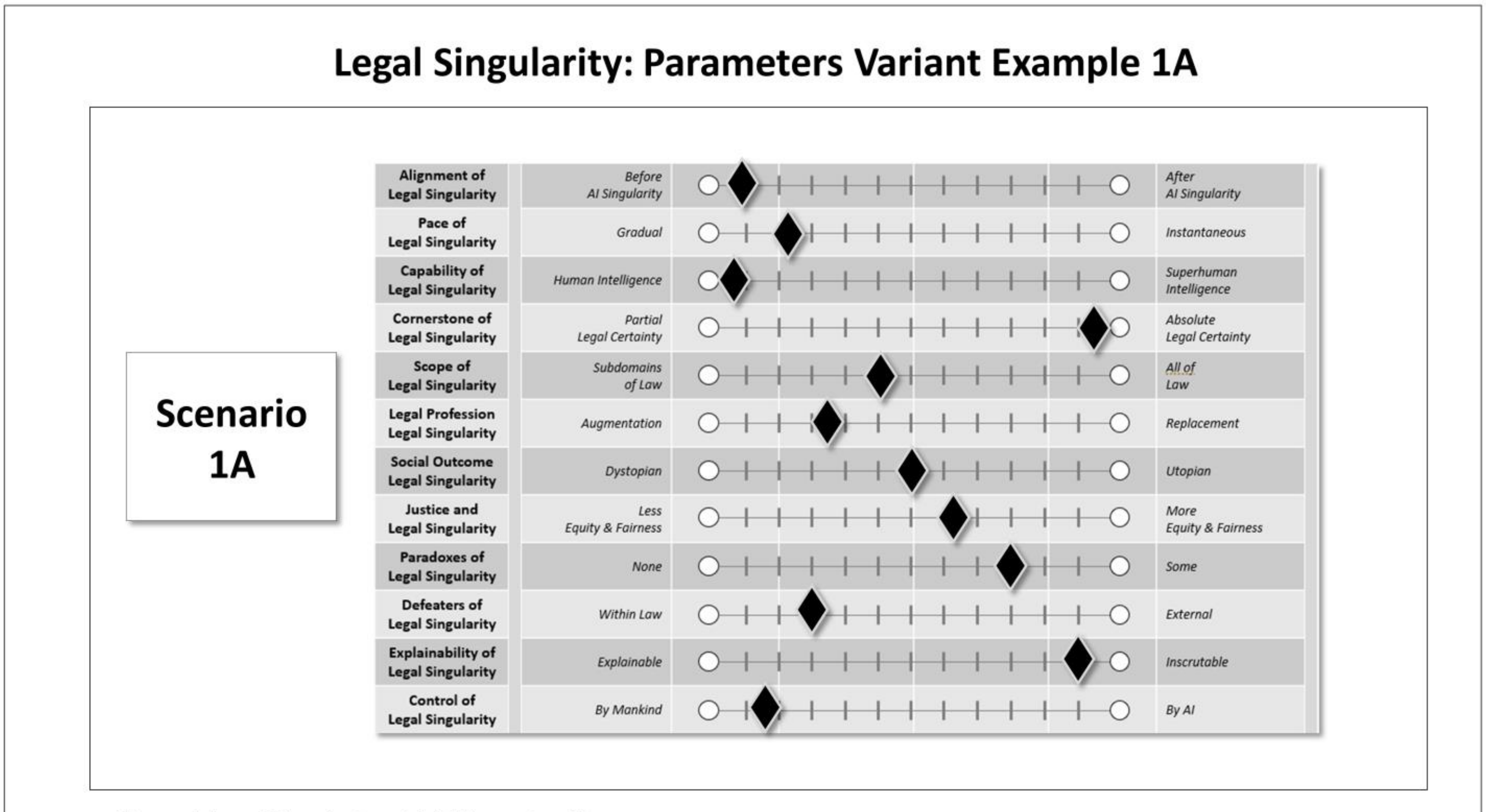

*Figure 4: Legal Singularity – Multidimensionality*

**Figure B-5**

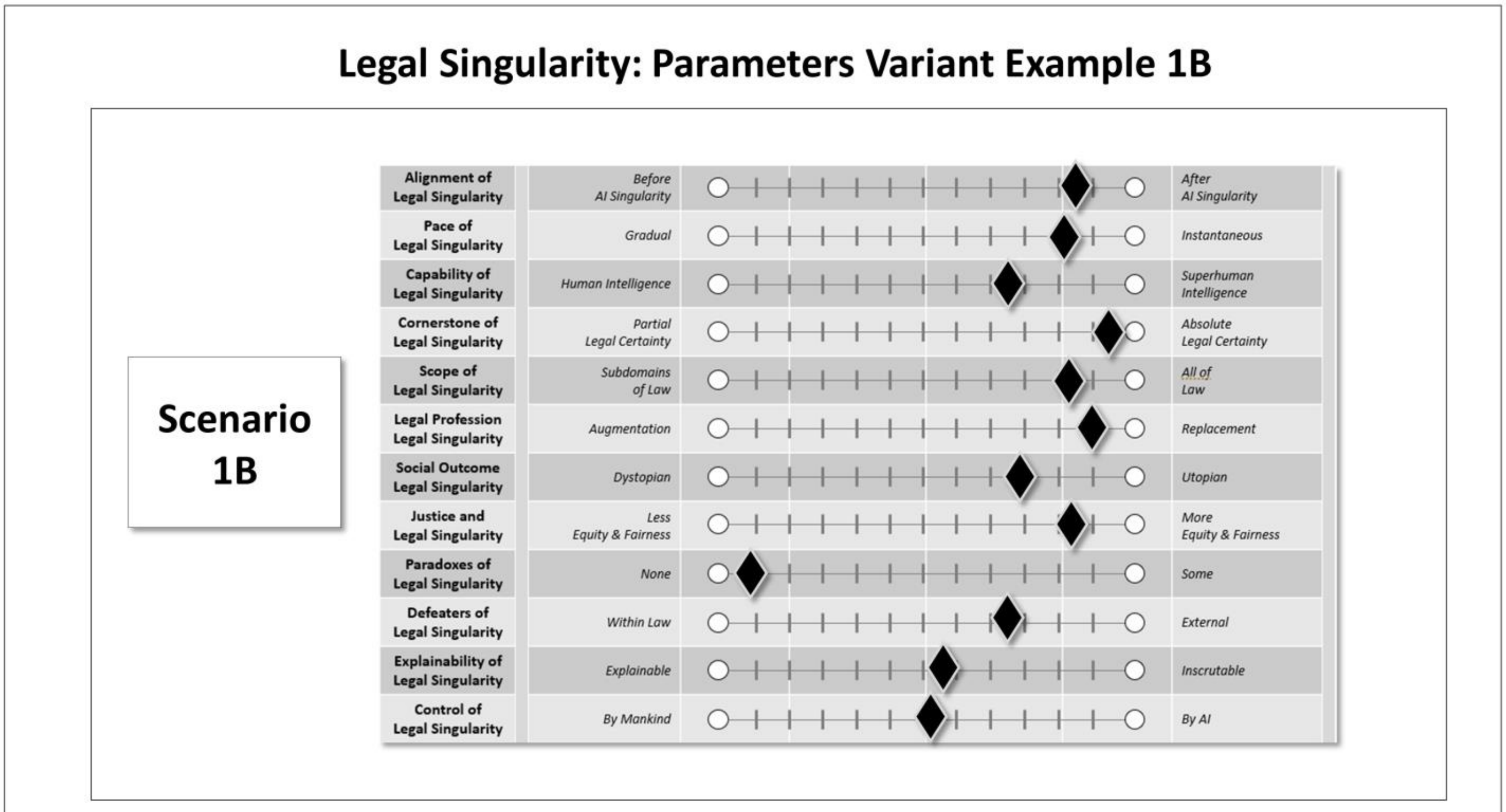

*Figure 5: Legal Singularity – Multidimensionality*

*Source Author: Dr. Lance B. Eliot*

V1.3